\documentstyle[aps]{revtex}

\begin{document}
\title{ Energy spectrum of a 2D Dirac oscillator in the presence of a constant magnetic field}

\author{V\'{\i}ctor M. Villalba\footnote{e-mail villalba@ivic.ve,
villalba@th-physik.uni-frankfurt.de
}}
\address{
Centro de F\'{\i}sica, Instituto Venezolano 
de Investigaciones Cient\'{\i}ficas IVIC, Apdo. 21827, Caracas 1020-A
Venezuela}
\author{Al\'{\i} A. Rinc\'on Maggiolo\footnote{e-mail alrincon@solidos.ciens.luz.ve}}
\address{Departamento de F\'{\i}sica, Universidad del Zulia. Maracaibo,
Venezuela}

\maketitle
\begin{abstract}
In this paper we obtain exact solutions of a 2D relativistic Dirac oscillator in the
presence of a constant magnetic field. We compute the energy spectrum and
discuss its dependence on the spin and magnetic field strength. 
\end{abstract}


\section{Introduction}

During the last years the study of electron in two-dimensional systems has
become a subject of \ active research. The rapid advances in
microfabrication technology has made possible to confine laterally two
dimensional electron systems. These quantum confined few-electron systems
are often referred to as artificial atoms where the potential of the nucleus
is replaced by an effective potential. A parabolic potential $V=\frac{1}{2}%
\omega r^{2}$ is often used as a realistic and at the same time
computationally convenient approximation. Despite its simplicity, parabolic
quantum dots appear to be a good approximation to complicated quantum dot
structures. \cite{Chakraborty}

The solution of the two-dimensional Schr\"{o}dinger equation for a single
electron in a homogeneous magnetic field ${\cal B}$ has been known after the
publication of the papers by Fock \cite{Fock} and Darwin \cite{Darwin} in
the 20's. The inclusion of a parabolic potential to this problem leads to
equations that can be solved without significative complications.

Despite the large body of papers discussing the confinement of electrons in
low dimensional structures, almost all of them tackle the problem in the
framework of the Schr\"{o}dinger equation. Among the difficulties when we
consider relativistic Dirac electrons with parabolic confinement we have
that we do not have at hand analytic solutions in terms of special functions
and we need to use semi-analytical and numerical methods in order to compute
the energy spectrum.

The relativistic extension of a parabolic confining potential can be done
with the of the Dirac oscillator \cite{Moshinsky,Benitez,Villalba}. Among
the advantages of this approach over the inclusion of Lorentz scalar
potential $\frac{1}{2}m\omega ^{2}r^{2}$, as confining potential, we have
the first one leads to a non-relativistic quadratic Hamiltonian in both
coordinates and momenta and therefore its solution can also be expressed in
terms of Laguerre or Hermite polynomials. The stability of the Dirac sea for
the Dirac oscillator Hamiltonian \cite{Martinez} also shows that this model
does not present difficulties related to the Klein's Paradox.

In the present article we solve the two-dimensional Dirac oscillator in the
presence of a constant magnetic field ${\cal B}$ perpendicular to the plane
where the electron is confined to move. \ The article is structured as
follows. In Sec. II, we solve the Dirac oscillator in polar coordinates. In
Sec. III, we analyze the energy spectrum of the oscillator.In Sec. IV, we discuss 
discuss our results and the behavior of the energy levels in the non relativistic limit. 
Along the paper we
adopt the natural units $\hbar =1,$ $c=1.$

\section{Solution of the Dirac oscillator}

The Dirac oscillator proposed by Moshinsky and Szczepaniak \cite{Moshinsky}
includes a type of interaction in the Dirac equation which, besides the
momentum, is also linear in the coordinates. The Dirac oscillator reduces,
in the non-relativistic limit, to a harmonic oscillator with a very strong
spin-orbit coupling term. Namely, the correction to the free Dirac equation 
\begin{equation}
i\frac{\partial \Psi _{c}}{\partial t}=(\beta {\bf \gamma p}+\beta m)\Psi
_{c}  \label{one}
\end{equation}
has the form: 
\begin{equation}
\label{dose}
{\bf p}\rightarrow {\bf p}-im\omega \beta {\bf r.}  \label{two}
\end{equation}
After substituting (\ref{two}) into (\ref{one}) we get an Hermitian operator
linear in both {\bf p} and {\bf r. }Recently, the Dirac oscillator has been
studied in spherical coordinates and its energy spectrum and the
corresponding eigenfunctions have been obtained \cite{Benitez}. Since we are
interested in studying the Dirac oscillator in a two-dimensional space, a
suitable system of coordinates for writing the harmonic interaction are the
polar $\rho $ and $\vartheta $ coordinates. In polar coordinates\ $%
(t,r,\vartheta )$ the metric tensor $g_{\alpha \beta }$ takes the form 
\begin{equation}
g_{\alpha \beta }=diag(-1,1,r^{2}),  \label{4}
\end{equation}
and the corresponding unitary vectors $\hat{e}_{\rho }\ \ $\ and $\hat{e}%
_{\vartheta }$ are 
\begin{equation}
\widehat{e}_{\rho }=\cos \theta \widehat{i}+sin\theta \widehat{j},\quad 
\widehat{e}_{\theta }=-sin\theta \widehat{i}+\cos \theta \widehat{j}
\end{equation}
In this case the radial component of the modified linear momentum takes the
form: $p_{\rho }-im\omega \beta \rho .$ The vector potential $\vec{A}$
associated with a constant magnetic field interaction is 
\begin{equation}
\vec{A}=\frac{{\cal B}\rho }{2}\hat{e}_{\vartheta }.  \label{5}
\end{equation}
The corresponding $\vec{{\cal B}}$ can be written in polar coordinates as
follows: 
\begin{equation}
\vec{{\cal B}}={\cal B}\hat{e}_{z}  \label{E}
\end{equation}
Since we are interested in computing the energy spectrum of a two
dimensional confined electron in the presence of a constant magnetic field,
in\ the present article, we analyze the solution of the 2+1 Dirac oscillator
in the presence of a constant magnetic field associated with the vector
potential (\ref{5}).

One begins by writing the Dirac equation (\ref{one}) in a given
representation of the gamma matrices. Since we are dealing with two
component spinors it is convenient to introduce the following representation
in terms of the Pauli matrices

\quad 
\begin{equation}
\beta \gamma _{1}=\sigma _{1},\ \beta \gamma _{2}=\sigma _{2},\ \beta
=\sigma _{3}  \label{rep}
\end{equation}

It is worth
mentioning that Dirac matrices expressed in the Cartesian tetrad gauge take
the form 
\begin{equation}
\widetilde{\gamma }^{\rho }=\vec{\gamma}\cdot \widehat{e}_{\rho }=\gamma
^{1}\cos \vartheta +\gamma ^{2}\sin \vartheta
\end{equation}
\begin{equation}
\widetilde{\gamma }^{\vartheta }=\vec{\gamma}\cdot \widehat{e}_{\vartheta
}=-\gamma ^{1}\sin \vartheta +\gamma ^{2}\cos \vartheta
\end{equation}
the similarity transformation $S(\vartheta )$ which reduces the Dirac
matrices $\widetilde{\gamma }^{\rho }$ and \ $\widetilde{\gamma }^{\theta }$
to $\gamma ^{1}$and $\ \gamma ^{2}$ is 
\begin{equation}
S(\vartheta )=\cos \frac{\vartheta }{2}-\gamma ^{1}\gamma ^{2}\sin \frac{%
\theta }{2}  \label{es}
\end{equation}
with 
\[
S^{-1}(\vartheta )\widetilde{\gamma }^{\rho }S(\vartheta )=\gamma ^{1},\quad
S^{-1}(\vartheta )\widetilde{\gamma }^{\vartheta }S(\vartheta )=\gamma ^{2}. 
\]
Using the matrix representation (\ref{rep}) and the similarity
transformation (\ref{es}), we obtain that the Dirac oscillator in the
presence of a constant magnetic field (\ref{E}) has the form

\begin{equation}
iE\Psi =H\Psi =\left[ \sigma ^{1}\partial _{\rho }+\sigma ^{2}(\frac{%
ik_{\vartheta }}{\rho }-m\omega \rho -\frac{e{\cal B}\rho }{2})+i\sigma
^{3}m\right] \Psi ,  \label{three}
\end{equation}
with, 
\begin{equation}
\Psi =\Psi _{0}(\rho )e^{i(k_{\vartheta }\vartheta -Et)},  \label{psi2}
\end{equation}
and 
\begin{equation}
\Psi _{0}=\pmatrix{\Psi _1 \cr \Psi _2\cr },
\end{equation}
where the spinor $\Psi $ is expressed in the (rotating) diagonal gauge. It
is related to the Cartesian (fixed) spinor by means of the transformation $%
S(\rho ,\vartheta )$\cite{Shishkin} 
\begin{equation}
\Psi =\sqrt{\rho }S(\vartheta )^{-1}\Psi _{c}  \label{ese}
\end{equation}
with 
\begin{equation}
S(\vartheta )\Psi _{d}=\Psi _{c}
\end{equation}
The matrix transformation $S(\vartheta )$ can be written as follows 
\begin{equation}
S(\vartheta )=\exp (-\frac{\vartheta }{2}\gamma ^{1}\gamma ^{2})=\exp (-i%
\frac{\vartheta }{2}\sigma _{3})
\end{equation}
The $\sqrt{\rho }$ factor has been introduced in (\ref{ese}) in order to
eliminate the term $\frac{S(\vartheta )\partial _{\vartheta }S(\vartheta
)^{-1}}{\rho }$. Noticing that $S(\vartheta )$ satisfies the relation 
\begin{equation}
S(\vartheta +2\pi )=-S(\vartheta )
\end{equation}
we obtain, 
\begin{equation}
\Psi (\vartheta +2\pi )=-\Psi (\vartheta )
\end{equation}
so we have $k_{\vartheta }=N+1/2$, where $N$ is an integer number. The
rotating Dirac spinor $\Psi _{d}$ can be written in terms of \ $\Psi _{c}$
as: 
\begin{equation}
\Psi _{d}=\left( 
\begin{array}{c}
e^{i\vartheta /2}\Psi _{1c} \\ 
e^{-i\vartheta /2}\Psi _{2c}
\end{array}
\right) ,\quad {\rm with\quad }\Psi _{c}=\left( 
\begin{array}{c}
\Psi _{1c} \\ 
\Psi _{2c}
\end{array}
\right) .
\end{equation}
We can label the quantum states of the Dirac oscillator in terms of \
eigenstates of the parity operator. In fact, analogously to the three
dimensional Dirac oscillator \cite{Benitez} we have that the parity operator
commutes with the Hamiltonian (\ref{three}) 
\begin{equation}
\lbrack P,H]_{-}=0
\end{equation}
where the parity operator acts on the Dirac spinor in the Cartesian gauge as
follows: 
\begin{equation}
P\Psi _{c}({\bf r})=\beta \Psi _{c}(-{\bf r})=\sigma _{3}\Psi (-{\bf r})
\end{equation}
In cylindrical coordinates, the reflection respecting to the origin ${\bf r}%
\rightarrow {\bf -r}$ is obtained via the rotation 
\begin{equation}
\rho \rightarrow \rho {\bf \quad }\vartheta \rightarrow \vartheta +\pi
\end{equation}
Taking into account the relation (\ref{ese}) between $\Psi $ and $\Psi _{c}$
as well as (\ref{psi2}) we readily obtain 
\begin{equation}
\Psi _{c}({\bf r})=\left( 
\begin{array}{c}
e^{i(k_{\vartheta }-1/2)\vartheta }\Psi _{1}(\rho ) \\ 
e^{i(k_{\vartheta }+1/2)\vartheta }\Psi _{2}(\rho )
\end{array}
\right)
\end{equation}
and consequently 
\begin{equation}
P\Psi _{c}({\bf r})=(-1)^{k_{\vartheta }-1/2}\Psi _{c}({\bf r})
\end{equation}
therefore, the parity of the energy eigenfunctions is given by $%
(-1)^{k_{\vartheta }-1/2}$. Reminding that the total angular momentum $J$
expressed in the Cartesian gauge takes the form 
\begin{equation}
J=-i(\frac{\partial }{\partial \vartheta }+\frac{\gamma ^{1}\gamma ^{2}}{2}),
\end{equation}
and taking into account the representation (\ref{rep}) for the Dirac
matrices we have that 
\begin{equation}
J\Psi _{c}({\bf r})=(k_{\vartheta }-\frac{1}{2}\sigma ^{3})\Psi _{c}({\bf r})
\end{equation}
then, the eigenvalues of the total angular momentum are 
\begin{equation}
j=k_{\vartheta }\mp \frac{1}{2}  \label{jota}
\end{equation}
.

Using the representation (\ref{rep}), the spinor equation (\ref{three}) can
be written as system of two first order coupled differential equations, 
\begin{equation}
i(E-m)\Psi _{1}(\rho )=(\frac{d}{d\rho }+\frac{k_{\vartheta }}{\rho }-\rho m%
\bar{\omega})\Psi _{2}(\rho ),  \label{uno}
\end{equation}
\begin{equation}
i(E+m)\Psi _{2}(\rho )=(\frac{d}{d\rho }-\frac{k_{\vartheta }}{\rho }+\rho m%
\bar{\omega})\Psi _{1}(\rho ),  \label{dos}
\end{equation}
where the frequency $\bar{\omega}$ can be written in terms of the Larmor
frequency as follows: 
\begin{equation}
\bar{\omega}=\omega +s\omega _{L}=\omega +\frac{e{\cal B}}{2m}
\end{equation}
substituting (\ref{dos}) into (\ref{uno}) and vice-versa we arrive at

\begin{equation}
\left[ \frac{d^{2}}{d\rho ^{2}}-\frac{(k_{\vartheta })(k_{\vartheta }\mp 1)%
}{\rho ^{2}}+m\bar{\omega}(2k_{\vartheta }\pm 1)-m^{2}\bar{\omega}^{2}\rho
^{2}+(E^{2}-m^{2})\right] \pmatrix{\Psi _1 \cr \Psi _2\cr }=0  \label{sist}
\end{equation}
It is not difficult to see that the solution of the second order equation (%
\ref{sist}) for $\Psi _{1}$ can be expressed in terms of associated Laguerre
polynomials $L_{k}^{s}(x)$\cite{Kamke,Abramowitz} as follows 
\begin{equation}
\Psi _{1}=c_{1}\exp (-x/2)x^{\frac{1}{2}(1/2+\mu )}L_{n}^{\mu }(x)
\label{Psi}
\end{equation}
where we have made the change of variables 
\begin{equation}
x=m\bar{\omega}\rho ^{2}
\end{equation}
the parameter $\mu $ is 
\begin{equation}
\mu =\pm (k_{\vartheta }-1/2)
\end{equation}
and the natural number $n$ satisfies the relation 
\begin{equation}
\frac{E^{2}-m^{2}}{m\bar{\omega}}+(1\mp 1)(2k_{\vartheta }-1)=4n
\label{ene}
\end{equation}
Since the function $\Psi _{1}$ must be regular at the origin, we obtain that
the sign of $\mu $ in (\ref{Psi}) is determined by the sign of $k_{\vartheta
}.$ In fact, for $k_{\vartheta }>0$ we have that $\Psi _{1}$ reads 
\begin{equation}
\Psi _{1}=c_{1}\exp (-x/2)x^{k_{\vartheta }/2}L_{n}^{k_{\vartheta }-1/2}(x)
\label{sol}
\end{equation}
substituting (\ref{sol}) into (\ref{dos}) we arrive at 
\begin{equation}
\Psi _{2}=2ic_{1}\frac{(m\bar{\omega})^{1/2}}{E+m}\exp
(-x/2)x^{(k_{\vartheta }+1)/2}L_{n-1}^{k_{\vartheta }+1/2}(x)  \label{sol2}
\end{equation}
where $c_{1}$ is an arbitrary constant.

\noindent Analogously, we obtain that the regular solutions for $%
k_{\vartheta }<0$ are, 
\begin{equation}
\Psi _{2}=c_{2}\exp (-x/2)x^{-k_{\vartheta }/2}L_{n}^{-k_{\vartheta
}-1/2}(x)
\end{equation}
\begin{equation}
\Psi _{1}=2ic_{2}\frac{(m\bar{\omega})^{1/2}}{E+m}\exp
(-x/2)x^{(1-k_{\vartheta })/2}L_{n}^{1/2-k_{\vartheta }}(x)
\end{equation}
where $c_{1}$ is a normalization constant The expression (\ref{ene}) can be
rewritten as follows 
\begin{equation}
E^{2}-m^{2}=4\left[ n-\Theta (-k_{\vartheta })(k_{\vartheta }-1/2)\right]
(m\omega +\frac{e{\cal B}}{2})  \label{spec}
\end{equation}
where $\Theta (x)$ is the Heaviside step function.

\section{Study of the energy spectrum}

From the relation (\ref{spec}) it is clear that the energy spectrum of the
2+1 Dirac oscillator depends on the value of $k_{\vartheta}$. Notice that for positive
values of $k_{\vartheta }$ the energy of the system has the form 
\begin{equation}
E^{2}=m^{2}+4n(m\omega + \frac{e{\cal B}}{2})
\end{equation}
For $k_{\vartheta }<0$ we observe that, the
states with $(n\pm l,$ $k_{\vartheta }-1/2\pm l),$ where $l$ is an integer
number, have the same energy. In this direction there are some differences
with the spherical Dirac oscillator \cite{Benitez}. Despite in both cases
bound states are obtained, for the 2+1 Dirac oscillator the energy spectrum
presents extra degeneracies only for negative values of $k_{\vartheta }.$
In order to get a deeper understanding of the dependence of the energy
spectrum on the spin we can take the nonrelativistic limit of the Dirac
equation (\ref{three}). In order to do that, it is advisable to work with
Eq. (\ref{sist}). The Galilean limit is obtained by setting $E=m+\varepsilon 
$, and considering $\varepsilon <<m$. Taking into account that the first two
terms in Eq.(\ref{sist}) are associated with the operator $P^{2}$, we obtain
in the nonrelativistic limit 
\begin{equation}
\frac{P^{2}}{2m}-(\omega +\frac{e{\cal B}}{2m})(k_{\vartheta }\pm \frac{1}{%
2})+(\omega +\frac{e{\cal B}}{2m})^{2}\frac{m^{2}\rho ^{2}}{2}=\varepsilon
\label{nonr}
\end{equation}
notice that Eq. (\ref{nonr}) corresponds to the Schr\"{o}dinger Hamiltonian
of a modified harmonic oscillator with an additional spin dependent term
given by $-\bar{\omega}(k_{\vartheta }\pm \frac{1}{2})$. This contribution
is proportional to the frequency of the oscillator plus the Larmor frequency.

The energy spectrum of the 2+1 Dirac oscillator can be computed using the
standard derivation followed by Moshinsky and Szczepaniak \cite
{Moshinsky,Lange}. In fact, using the Bjorken and Drell \cite{Bjorken}
representation for the Dirac matrices, the Dirac oscillator coupled to an
electromagnetic field reads

\begin{equation}
(E-m)\Psi _{1}=\sigma \cdot \lbrack ({\bf p}-e{\bf A})+im\omega {\bf r]}\Psi
_{2}
\end{equation}
\begin{equation}
(E+m)\Psi _{2}=\sigma \cdot \lbrack ({\bf p}-e{\bf A})-im\omega {\bf r]}\Psi
_{1}
\end{equation}
taking into account that the vector potential for a constant magnetic field $%
{\cal B}$ is given by (\ref{5}) and we are working in 2+1 dimensions, we
obtain 
\begin{equation}
(E^{2}-m^{2})\Psi =\left( (p^{2}+m^{2}\bar{\omega}^{2}\rho ^{2})-2m\bar{%
\omega}-4m\bar{\omega}L_{z}S_{z}\right) \Psi  \label{ene1}
\end{equation}
where 
\begin{equation}
L_{z}=({\bf r}\times {\bf p)}_{z}\quad S_{z}=\frac{\sigma _{3}}{2}
\end{equation}
finally, recalling the energy spectrum of a non relativistic harmonic
oscillator we readily obtain that expression (\ref{ene1}) takes the form 
\begin{equation}
\left( E^{2}-m^{2}\right) \Psi =\left( 2m\bar{\omega}(2n+(k_{\vartheta }\mp 
\frac{1}{2})+1)-2m\bar{\omega}-2m\bar{\omega}(k_{\vartheta }\mp \frac{1}{2}%
)\right) \Psi
\end{equation}
\begin{equation}
E^{2}-m^{2}=\left[ 4n+2(-2k_{\vartheta }+1)\Theta (-k_{\vartheta })\right]
m\bar{\omega}  \label{spectrum}
\end{equation}
result that coincides with expression (\ref{spec}).

We can label the eigenstates of the parity operator in terms of the total
angular momentum $\ j.$ If we define 
\begin{eqnarray}
\epsilon &=&-1\textrm{ if parity is }(-1)^{j} \\
\epsilon &=&1\textrm{ if parity is }(-1)^{j-1}  \nonumber
\end{eqnarray}
in both cases we have $k_{\vartheta }=j+\frac{\epsilon }{2}$, therefore we
have that the energy spectrum (\ref{spectrum}) can be
written in terms of the parity $\epsilon $ as follows 
\begin{equation}
E^{2}-m^{2}=\left[ 4n+2(2k_{\vartheta }\epsilon +1)\Theta (\epsilon
k_{\vartheta })\right] m\bar{\omega}
\end{equation}
From (\ref{spectrum}) we can see that the energy spectrum of the Dirac
oscillator in the presence of a constant magnetic field presents a
dependence on ${\cal B}$ that differs from that obtained with the help of
the Pauli equation with a harmonic potential. \cite{Akhiezer,victor}. In this
case the spin couples to the magnetic field via the term $-\frac{e{\cal B}}{%
2m},$ and there is a contribution of ${\cal B}$ on the energy for positive
as well as for negative spin orientations.

From (\ref{spectrum}) it is straightforward to obtain the energy spectrum $S 
$ states. For $k_{\vartheta}>0$ we have 
\begin{equation}
\label{ener1}
E^{2}-m^{2}=4nm\bar{\omega}=4n(m\omega +\frac{e{\cal B}}{2})  \label{mas}
\end{equation}
on the other hand, for $k_{\vartheta}<0$, \ the energy spectrum satisfies the relation 
\begin{equation}
\label{ener2}
E^{2}-m^{2}=4(n+\frac{1}{2}-k_{\vartheta})(m\omega +\frac{e{\cal B}}{2})  \label{menos}
\end{equation}
from (\ref{mas}) and (\ref{menos}) we observe that the energy spectrum of
the Dirac oscillator in the presence of a constant ${\cal B}$ depends on the
spin orientation. For $k_{\vartheta}>0$ the energy of $S$ ground state satisfies $%
E=m$. Conversely, for $k_{\vartheta}<0$ the energy $E$ satisfies the relation \ $%
E^{2}=m^{2}+4(m\omega +\frac{e{\cal B}}{2})$ and consequently depends on the
magnetic field strength ${\cal B}$.

\section{Discussion of the results}

The energy spectrum of a non relativistic two-dimensional electron confined by
a parabolic potential in the presence a of static magnetic field $\cal B$ is
\cite{Fock,Darwin}

\begin{equation}
\label{energie}
E_{n'}=(2n'+\mid l \mid +1)\hbar \Omega -\frac{1}{2}\hbar \omega_{c}l
\end{equation}
where $n'=0,1,2..$, and  $l=0, \mp 1, \mp 2,...$. 
The renormalized oscillator 
frequency $\Omega$ reads 
\begin{equation}
\Omega=\sqrt{\omega^2+\frac{\omega_{c}^2}{4}}
\end{equation}
The expression (\ref{energie}) does not contain the spin field coupling term
appearing in the Pauli equation: 
\begin{equation}
g^{*}\mu_{B}\vec{\cal B}\cdot   {\vec{S}}=\pm \frac{g^{*}e}{4m}{\cal B}
\end{equation}
that, for the value $g^{*}=2$, splits the energy levels by a factor $\pm \omega_{c}$ 

In order to better understand the energy spectrum (\ref{ener1}), (\ref{ener2}), we proceed to 
analyze the asymptotic limits of the non minimal coupling introduced in Eq.(\ref{dose}).  

The weak oscillator limit can be obtained after taking  the limit  
$\omega \rightarrow 0$. The energy spectrum  (\ref{ener1}), (\ref{ener2})
reduces to
\begin{equation}
\label{esp}
E^2=m^2+2\left[ n-\Theta (-k_{\vartheta })(k_{\vartheta }-1/2)\right]e{\cal B}
\end{equation}
which are the relativistic Landau levels obtained when one solves the Dirac equation. 
The non relativistic limit can also be obtained in a straightforward way from (\ref{esp}) 
giving as result the expression (\ref{energie}). The energy spectrum in weak magnetic field limit 
presents a behavior that is not observed in the non relativistic oscillator case. When we eliminate
the magnetic field strength  $\cal B$ the energy spectrum remains degenenerate but depends on the value of $k_{\vartheta}$ therefore, in this limit the Dirac oscillator does not seem to be a suitable approximation to the problem of an electron confined in a quantum dot 
\cite{Chakraborty,Ferry}.  

\section{Concluding remarks}

The presence of a constant magnetic field ${\cal B}$ \ (\ref{E}) does not
break up the symmetry of the vacuum \cite{Martinez} and therefore we do not
have mixing of positive and negative energy solutions. The confining Dirac
oscillator potential gives as a result a spin orientation  dependent
energy spectrum for the electron in a the presence of a constant magnetic
field. There are extra degeneracies of the quantum states with the same
energy when $k_{\vartheta }<0.$The non relativistic limit of the model
presented in this paper shows that ${\cal B}$ modifies the oscillator
frequency $\omega $ via the Larmor term $\omega _{L}=\frac{e{\cal B}}{2m}.$. 
The nature of of the coupling associated with the Dirac oscillator does not
not permit one to use this model in the study of an electron  
confined by a quantum dot. 

\vspace{0.5cm}
\acknowledgements

\noindent This work was supported by CONICIT under project 96000061.


\begin{references}

\bibitem{Chakraborty}  T. Chakraborty, Comments Cond Mat. Phys. {\bf 16} 
(1992) 35

\bibitem{Fock}  V. Fock, Z. Physik {\bf 64} (1928) 629

\bibitem{Darwin}  C. G. Darwin, Proc. Cambridge Phil. Soc. {\bf 27} 
(1930) 86.

\bibitem{Moshinsky}  M. Moshinsky and A. Szczepaniak, J. Phys. A {\bf 22, }
(1989) L817.

\bibitem{Benitez}  J. Ben\'{\i}tez, R. P. Mart\'{\i}nez y Romero, H. N. N\'{u}%
\~{n}ez-Y\'{e}pez, and A. L. Salas-Brito, Phys. Rev. Lett. {\bf 64}
(1990) 1643.

\bibitem{Villalba}  V. M. Villalba, Phys. Rev. A, {\bf 49}, (1994) 586.

\bibitem{Martinez}  R. P. Mart\'{\i}nez, Y. Romero, M. Moreno, and A.
Zentella, Phys. Rev. D. {\bf 43, } (1991) 2036.

\bibitem{Shishkin}  G. V. Shishkin and V. M. Villalba, J. Math. Phys{\it .}, 
{\bf 30}, (1989) 2132.

\bibitem{Kamke}  E. Kamke, \textit{Differential\-gleichun\-gen L\"{o}sungs\-metho\-den
und L\"{o}sun\-gen, Band 1: Gew\"ohnliche Differentialgleichingen} (Akademische 
Verlagsgesellchaft Geest und Portig KG, Leipzig, 1951).

\bibitem{Abramowitz}  M Abramowitz and I. Stegun, {\it Handbook of
Mathematical Functions,} Natl. Bur. Stand. Appl. Math. Ser. No. 55 U.S.
GPO, Washington, D.C., 1965.

\bibitem{Lange}  O. L. de Lange and R. E. Raab {\it Operator Methods in
Quantum Mechanics} New York, Oxford U. P. 1991.

\bibitem{Bjorken}  D. Bjorken and S. Drell, {\it Relativistic Quantum
Mechanics}, McGraw Hill, New York,  1964.

\bibitem{Akhiezer}  A. I. Akhiezer an V. B. Berestetskii, {\it Quantum
electrodynamics} Moscow, Nauka, 1969.

\bibitem{victor}  V. M. Villalba and R. Pino, Phys. Lett A. {\bf 238}
(1998) 49.

\bibitem{Ferry} D. K. Ferry and S. M. Goodnick {\it Transport in Nanostructures}
(Cambridge University Press, Cambridge 1999).

\end{references}
\end{document}